\documentstyle[prl,aps]{revtex} 

\begin{document}
\draft
\title{ An exactly solvable model of the persistent currents
in presence \\
of hard-core interaction in a one dimensional mesoscopic ring}

\author{You-Quan Li$^{1}$ and Zhong-Shui Ma$^{2}$ }

\address{
${}^{1}$Zhejiang Institute of Modern Physics,
Zhejiang University, Hangzhou 310027, China \\
${}^{2}$Advanced Research Center, Zhongshan University,
Guangzhou 510275, China
}
\date{Received December 18, 1995 }
\maketitle
\begin{abstract}
A simplest model of the persistent current in presence of many-body
interaction in a mesoscopic  ring is presented.
The Bethe ansatz approach is used to find
exact solutions. Both cases in the absence of impurity and in the presence of
impurity are solved and compared. It is indicated that both impurity
and hard-core interaction do not affect the result of persistent current
explicitly if there is a large number of electrons in the ring.
\end{abstract}



It is well known that quantum coherence  plays a central role in
mesoscopic physics. The persistent current on mesoscopic rings
threaded by a magnetic flux is a particular sensitive probe of
such coherence. The existence of persistent currents on  mesoscopic
rings was predicted early in [1-3], but it had not absorbed much
attention for a long time due to the technological reasons. Owning to
the advances in nanotechnology, more research interests were involved
in both  theoretical [4,5] and  experimental [5-7] aspects. Among these
theoretical researches, \cite{Ch1} studied the model that the
electrons move freely in an ideal ring. The effect of a periodic
potential was considered in the framework of hopping
model \cite{Ch2}. The influence of impurity was calculated mumerically
\cite{Mon} and analytically in the framework of the grand canonical
ensemble \cite{Ch3} The consideration of a screened Coulomb interaction
was discussed \cite{Amb} via the Hartree-Fock approximation, and the
long-range part of the Coulomb interaction was taken into account
\cite{Kop} by means of the energy-density functional. There are
also other discussions, such as a combination of impurity-
and Coulomb- perturbation \cite{Eck,Mul}, hopping models with
a short-range two-body interaction\cite{Kus,Abr}, and the model in terms of
the Luttinger liquid \cite{Los} or that from the Bogliubov-de
Gennes equation \cite{Wang} etc.. However, as far as we were aware,
there is not a thorough discussion about this subject by
an exactly solvable many-body model. Present letter is an attempt
toward this destination. As a result of our discussion, it is
indispensable that the wave-function must be piecewisely continuous
functions if one consider $N$ electrons in a one dimensional ring.
As long as the circumference of the ring is supposed to be the mesoscopic
scale such that the inelastic scattering is considered not to happen,
it is applicable to use the Bethe ansatz approach to solve the
present problem. In  the following, we discuss  the problems of
$N$ particles and that of $N$ particles with a single impurity on a
mesoscopic ring in the presence of magnetic flux respectively.
The results of both cases are compared.

The model Hamiltonian for the system is supposed to be
a $N$-body  problem with delta-function interactions on a ring.
For clarity and convenience,
we convert the nature of the ring to be a system on a line with periodicity
of $L$. The Hamiltonian reads
\begin{equation}
H = - \frac{\hbar^{2}}{2m}\sum^N_{i=1}(\frac{\partial }{\partial x_{i}} - i\frac{2\pi}{L}
\frac{\phi}{\phi_{0}})^{2} +  u\sum^N_{j>i=1}\delta(x_{i} - x_{j} - nL),
\label{eq:a}\end{equation}
where $u$ stands for the coupling constant;
$n$, $\phi_{0}$ and $\phi$ stand for integers, the flux quantum $hc/e$
and the total magnetic flux threading the ring, respectively.

It is easy to show that if $\psi'$ is an eigenfunction of the Schr\"{o}dinger
equation in the absence of magnetic flux, then
$\psi(x_{1}, x_{2}, \cdots )
=\exp \left( \displaystyle\frac{2\pi i }{L}
\frac{\phi}{\phi_0 }
(x_1 + x_2 + \cdots ) \right) \psi'(x_{1}, x_{2}, \cdots )$
satisfies the Schr\"{o}dinger equation in the presence of magnetic flux.
Moreover, if $\psi'$ satisfies the periodic condition then $\psi$ does
not satisfy periodic condition but an additional torque
(non-integrable phase)
$\exp( i 2\pi\phi/\phi_{0})$ comes about. However the wave-function of
the real system, being on a ring, is required to be single valued. This implies that
the periodic condition must be imposed upon the $\psi$ i.e. .
$\psi (x_{1}, \cdots, x_{i} + L, \cdots )=\psi(x_{1},\cdots, x_{i},\cdots)$.
Apparently, it is guaranteed as long as the wave-function
$\psi' $ is assumed to satisfy a twisted-periodic condition\cite{Sri,Kor}
instead of the usual periodic one. Therefore, the calculation of the
energy spectrum of a
Schr\"{o}dinger equation in the presence
of a magnetic flux with periodic condition is equivalent to that
in the absence
of the flux with, whereas, a twisted-periodic condition, i.e.
\begin{equation}
\psi (x_{1}, \cdots, x_{i} + L, \cdots ) = \exp (i2\pi \phi /\phi_{0})
\psi (x_{1}, \cdots, x_{i}, \cdots )
\label{eq:b}\end{equation}
Because of the above mentioned equivalence, we only needs to solve the eigenequation
of Hamiltonian (\ref{eq:a}) by taking the flux as null while taking account of
the twisted periodic condition (\ref{eq:b}). We emphasize
that the characteristic  length
$L$  of the circumference is within the mesoscopic scale such that the
inelastic scattering does not
happen. This suggests the validity of Bethe ansatz approach for present
problem because the ansatz embodies the property of non-diffractive 
(i.e.  elastic in one-dimension) scattering.
Apart from the consideration of the twisted periodic condition, the whole
course of solving the problem is well known \cite{Yang,Li}. So we do not
exhibit the details but only give the main results that are tightly related
to our following discussion. The energy spectrum of Bethe ansatz solution is
given by
$E = \displaystyle{\frac{\hbar^{2}}{2m}\sum^N_{j=1}k_{j}^{2} }$, where
$\{k_{j}\} $ are roots of the following transcendental equations
\begin{equation}
k_{j} = (I_{j}+ \frac{\phi}{\phi_{0}} ) (\frac{2\pi }{ L})
- \sum^N_{l=1}\tan^{-1}\left(\frac{ k_{j} - k_{l}}{u}\right).
\label{eq:c}\end{equation}
 The $\{ I_{j} \}$
play the role of quantum numbers, and take integer value or half-integer
value accordingly as  $N=$ is odd or even. As the Bethe
ansatz wave function becomes null as long as two $k$'s take the same value,
the system has fermionic property even for N-bosons. Thus the quantum numbers
of ground state are
$
\{ -\frac{N-1}{2}, - \frac{N-1}{2} + 1, \cdots, \frac{N-1}{2} -1,
\frac{N-1}{2} \}
$

If let $k'_{j} = k_{j}
- \displaystyle{
\frac{2\pi}{L}
(\frac{\phi}{\phi_0 })
}$,
(\ref{eq:c}) becomes the usual secular equation for $\{k'_{j} \}$
\begin{equation}
k'_{j} = I_{j}
(\frac{2\pi }{ L})
- \sum^N_{l=1}\tan^{-1}\left(\frac{ k'_{j} - k'_{l}}{u}\right)
\label{eq:d}\end{equation}
in which the magnetic flux is not involved (and it agrees with the result
of \cite{Yang}). So the roots of (\ref{eq:d}) are
independent of the flux $\phi$, which is useful in deriving the following
(\ref{eq:f}). In terms of these roots we can write out
the energy of the system as follows:
\begin{equation}
E = \frac{\hbar^{2}}{2m}\sum^N_{j=1}
\left[ k'_{j} + \frac{2\pi}{L}(\frac{\phi}{\phi_{0}})
\right]^2 .
\label{eq:e}\end{equation}
It is different from the result of the free electron model where 
$k'_{j}$ is an integer instead of the roots of the secular equation (\ref{eq:d}).
The knowledge of the full energy spectrum enables us to calculate the
persistent current  of the system by using the formula  in \cite{BYang}.
For a single ring at zero temperature,
the persistent current is calculated by
$ I(\phi )= - \displaystyle{c \frac{d E}{d \phi }}$,
explicitly,
\begin{eqnarray}
I(\phi)& = & - \frac{ e\hbar}{ m L }
\sum^N_{j=1}
\left[ k'_{j} + \frac{2 \pi}{L}
( \frac{\phi}{\phi_{0}})
\right] \nonumber \\
& = & -\frac{N}{L} \frac{ eh}{mL}(n + \frac{\phi}{\phi_{0}} ).
\label{eq:f}\end{eqnarray}
In deriving the second line of (\ref{eq:f}), (\ref{eq:d}) has been used.
As a result, $n:=\frac{1}{N}\sum_{j=1}^N I_j $ is an integer regardless
of $ N$  being even or odd.
For $\phi = 0 $, the ground state corresponds to $n=0$, and the first excited
state corresponds to $ n= \pm 1 $. However, for $\phi = \pm \phi_0 $,
the ground state corresponds to $ n = \mp 1$. Similarly, one can find that
the current $I(\phi)$ is a sawtooth-type function with periodicity of
$ \phi_0 $.

Now let us turn to the  discussion  of the situation  in the presence
of a single impurity, the Hamiltonian  is given by
\begin{eqnarray}
H = - \frac{\hbar^{2}}{2m}\sum^N_{i=1}
\left[
\frac{\partial }
{\partial  x_{i}}
- i\frac{2 \pi}{L}( \frac{\phi}{\phi_{0} } )
\right]^{2}
- \frac{\hbar^{2}}{2m'}
\left[ \frac{\partial }{\partial  x'}
- i\frac{2 \pi}{L}(\frac{\phi}{\phi_{0}})
\right]^{2}  \nonumber\\
+ u\sum^N_{j>i=1}\delta(x_{i} - x_{j} - n L)
+ v\sum^N_{i=1}\delta(x_{i} - x' -n'L)
\label{eq:g}
\end{eqnarray}
where $x'$ stands for the coordinate of the impurity and $x_{i}$ for that of
the $i$th particle. It should be addressed again that the energy spectrum for
the given Hamiltonian is equivalent
to that  of the same Hamiltonian in the absence of magnetic flux
but with a twisted-periodic condition instead of the usual one.
Similar to our previous discussion in \cite{LiMa1},
 we take a scalar transformation
$ x' \rightarrow x_{0} =\sqrt{m'/m }x'$.
With this transformation, one can write the kinetic part
of both electrons and the impurity to a standard Laplace operator 
in an $(N+1)$ dimensional euclidean space with cartesian coordinates.
At the same time we would obtain the boundary condition arising from the
delta-function terms conveniently.
By using Gaussian integral theorem, one can derive the correct
discontinuity relations for the derivatives
of wave-function arising from  the delta-function in Hamiltonian (\ref{eq:g}).
The total energy is
$
E =\frac{\hbar^{2}}{2m}\sum^N_{j=1}k_{j}^{2} + \frac{\hbar^{2}}{2m'}\lambda^{2}
$
and the corresponding wave-function is obtained  to be
$
\psi(y_{0}, y_{1}, \cdots, y_{N} ) = e^{iK y_{0}}
\varphi(y_{1}, \cdots, y_{N} )
$,
where
$
K:=\sum^N_{j=1}k_{j} + \lambda $ and $y_{i} (i=0, 1, 2,\cdots, N)
$
are the coordinates of particles in the frame of reference of impurity.
The twisted periodic boundary condition leads to
$
K =  n'(\frac{2\pi}{L}) + (N+1)\frac{\phi}{\phi_0 }(\frac{2\pi}{L})
$.
The Bethe ansatz wave-function
$\varphi $ is a piecewisely continuous function defined on separate regions
which are specified by permutation group \cite{LiMa1} $S_{N+1}$, namely
\begin{equation}
\varphi_{\tau }(y) = \sum_{\sigma \in S_{N} } A(\sigma, \tau ) 
e^{i(\sigma k | y ) }, \,\, \tau \in S_{N+1}
\label{eq:h}\end{equation}
where $\sigma k $ stands for the image of a given 
$k:= (k_{1}, k_{2}, \cdots, k_{N})$
by a mapping of permutation
$\sigma \in S_{N}$ and the coefficients
$A(\sigma, \tau )$ are functionals on
$S_{N} \otimes S_{N+1}$. It is worthwhile to mention that the summation in
(\ref{eq:h}) runs over the permutation group $S_{N}$, whereas the
various regions on which the wave-function is defined are specified
by $S_{N+1}$. This differs from the Bethe-Yang ansatz \cite{Yang}.

All the coefficients A's in (\ref{eq:h}) are determined up to an overall scalar
factor. The coefficients among different regions are  related by the
following relations:
\begin{equation}
A(\sigma, \sigma_{i}\tau ) = \eta A(\sigma_{i}\sigma, \tau ), \,\,{\rm for}\,
i= 1, 2,\cdots, N-1.
\label{eq:i}\end{equation}
in which
$\eta =1 \, ( {\rm or} -1 )$
for totally symmetric (or totally anti-symmetric) state, and
\begin{equation}
A(\sigma, \sigma_{N}\tau ) 
= - \frac{v-i[(\sigma k)_{i}-\lambda /\mu ]}{v+i[(\sigma k)_{i}-\lambda /\mu ]}
A(\sigma, \tau ), 
\label{eq:j}\end{equation}
where $\mu = m'/m$; $\sigma_{N} \in S_{N+1}$ but $\not\in S_{N}$.
The relation of the coefficients on any one of the regions is given by
\begin{equation}
A(\sigma_{i}\sigma, \tau ) =
- \frac
{ u -i [(\sigma k )_{i} - (\sigma k )_{i+1} ] \eta }
{  u + i [(\sigma k )_{i} - (\sigma k )_{i+1} ]   }
A(\sigma, \tau ) .
\label{eq:k}\end{equation}
If $y=(y_{1},\cdots, y_{i},\cdots, y_{N})$
is a point on the region specified by
$\tau \in S_{N+1}$, as a result of periodicity, the coordinate
$y' =(y_{1},\cdots, y_{i} + L, \cdots, y_{N} )$
must correspond to a point on another region specified by
$\gamma\tau \in S_{N+1}$, where
$\gamma = \sigma_{N}\Delta $ and
$\Delta := \sigma_{N-1}\cdots \sigma_{2}\sigma_{1} \in S_{N} $.
Then the requirement of the twisted-periodic condition yields
$\varphi_{\gamma\tau }(y')=\exp( i 2\pi\phi/\phi_{0})\varphi_{\tau }(y)  $.
After writing out this relation  in terms of the Bethe ansatz 
solution (\ref{eq:h}), we find that the twisted periodic condition is guaranteed as long
as the  relation
$A(\sigma, \gamma\tau ) e^{ i(\sigma k )_{1} L } 
=\exp( i 2\pi\phi/\phi_{0}) A(\sigma, \tau )$
is imposed. By making use of (\ref{eq:j}), (\ref{eq:k}) and
(\ref{eq:i}) repeatedly, we obtain
\begin{equation}
e^{i(\sigma k)_{1}L} = (-1)^{N}e^{ i 2\pi\phi/\phi_{0} } \frac
{v - i[(\sigma k)_{1} - \lambda /\mu ]}
{v + i[(\sigma k)_{1} - \lambda /\mu ]}
\prod^{N}_{j=1}\frac
{u - i\eta [(\sigma k)_{1} - (\sigma k)_{j}]}
{u + i[(\sigma k)_{1} - (\sigma k)_{j}]}
\label{eq:l}\end{equation}
Taking the logarithm of (\ref{eq:l}), we have a system of coupled transcendental
equations,
\begin{equation}
k_{j} = (I_{j} + \frac{\phi} {\phi_0} )
(\frac{2\pi}{ L}) - \frac{2}{L}
\left[
\tan^{-1}\left(\frac{ k_{j} - \lambda /\mu }{v}\right)
+ \frac{ 1 + \eta }{2}
\sum^N_{l=1}\tan^{-1}\left(\frac{ k_{j} - k_{l} }{u}\right)\right]
\label{eq:m}\end{equation}
where  $I_{j}$ takes integer values or half integer
values respectively for $N$ being an even or odd number.
(\ref{eq:m}) is the secular
equation for the spectrum and the $ \{ I_{j} \}$ play the role of
quantum numbers. 
From (\ref{eq:h}) we learned that an eigenstate
related to $\{ k_{1}, k_{2},\cdots, k_{N} \}$
has the same eigenenergy as those related to
$\{ \lambda, k_{2},\cdots, k_{N} \}$, $\{ k_{1}, \lambda,\cdots, k_{N} \}$
etc. if $\mu =1 $. However, the $N+1$-fold degeneracy
is broken by the replacement of one of the particle with an impurity.

If let
$
k'_j = k_j - \frac{\phi}{\phi_0 }( \frac{2\pi}{L} )
$
and
$
\lambda' = \lambda - \mu \frac{\phi}{\phi_0 } (\frac{2\pi }{L}),
$
(\ref{eq:m})  becomes the same secular equation as in \cite{LiMa1}:

\begin{equation}
k'_j = I_j ( \frac{2\pi}{L} ) - \frac{2}{L}
\left[
\tan^{-1} \left(  \frac{ k'_j - \lambda' /\mu }{  v  } \right)
+ \frac{ 1 + \eta }{2}
\sum^N_{l=1} \tan^{-1} \left( \frac{ k'_j - k'_l }{ u } \right)
\right]
\label{eq:n}\end{equation}
in which the magnetic flux is not involved. Thus the roots of the equation
are independent of the flux $\phi $. In terms of these roots we can write
out the energy as follows
\begin{equation}
E = \frac{\hbar^2 }{ 2m } \sum^N_{j=1}
\left[
k'_j + \frac{\phi}{\phi_0 } (\frac{ 2 \pi }{ L })
\right]^2
+ \frac{\hbar^2}{ 2 m' }
\left[ \lambda' + \mu \frac{\phi }{ \phi_0 } (\frac{ 2 \pi }{ L })
\right]^2.
\label{eq:o}\end{equation}

The persistent current is formally calculated as
\begin{equation}
I(\phi ) =  -\frac{N+1}{L}\cdot\frac{ eh }{ mL }
( n + \frac{\phi }{\phi_0 })
\label{eq:p}\end{equation}
which is just the same result of (\ref{eq:f}) for $ N +1 $ electrons.
This result is of no surprise. As an equilibrium quantity, the persistent
current at zero temperature is obtained by taking summation of the currents
over all
states with energies below the Fermi level. Due to the hard-core interaction,
The Fermi statistics confers upon those states. The impurity plays a similar
role as an electron in occupying those lowest-lying single-particle levels
with the corresponding quantum number $\lambda'$, which is dependent on the mass
of impurity. For the originally occupied state with $k'_j $ for the $j$th
electron, the quantum number would change into what
associated with impurity occupying the  state.
The current for any occupation of $N$ electrons and  an impurity in Fermi sea
and at the Fermi surface is the same. However, electron beneath the Fermi
surface is raised to the Fermi surface or becomes excited state, this
increases the result in (\ref{eq:f}) by
$ -\frac{eh}{mL} \frac{1}{L} ( n + \frac{\phi}{\phi_0 } ) $.
We would emphasize that the effect of impurity scattering have not considered
here so that we have not dealt with the effect of localization in the
amplitude of persistent current which is given  by that of impurity
slowing down the moving electron.

Obviously, the assumption of the impurity having the same charge as the
electrons might be too special. If the impurity has a different electric
charge $q$, one can make the same discussion as we did in the above.
In the case, the twisted-periodic condition reads
\begin{eqnarray}                      
\psi (x_1, ..., x_i + L, ..., x_N, x') =
\exp (i2\pi \phi /\phi_0 ) \psi (x_1, ..., x_i, ..., x_N, x') \nonumber \\
\psi (x_1, ..., x_N, x'+ L) =
\exp (i2\pi\nu \phi /\phi_0 ) \psi (x_1, ..., x_i, ..., x_N, x')
\end{eqnarray}
where $\nu = q / e $. The formula for the persistent current becomes
\begin{equation}
I(\phi ) =  -\frac{N+1}{L}\cdot\frac{ eh }{ mL }
( n + \frac{N + \nu}{N+1}\frac{\phi }{\phi_0 })
\label{eq:r}\end{equation}
Clearly, the variance of electric charge of impurity from that of electrons
affects the persistent current only when the number of electrons is not
very large. Finally it needs to indicate that the formula (\ref{eq:p})
and (\ref{eq:r}) are not valid for the case of
$\mu = \infty$. As far as this case, the problem will become $N$ electrons
moving in a ring with a narrow ($\delta$-function type) junction because
the model Hamiltonian do not have translational invariance.

In summary, we have proposed an exactly solvable model in which the
interactions of both electron-electron and  electron-impurity
are taken into account. As a result, the persistent current in
a one dimensional mesoscopic
ring are exactly the same for either free electrons or electrons with
hard-core interaction, though the energy spectrums of them might be quite
different. On the other hand, from the above discussion we found that
a single impurity do not affect the result of persistent current at least
for the case of hard-core interaction if there is a large number of electrons
in the ring.

This work is supported by NSFC and NSF of Zhejiang province, partially supported
by CAS foundations and the foundation of Advanced Research Center of Zhongshan
University. The authors are grateful to the referee for his helpful comments
and to the Physical Society of Japan for financial support in publication.

\end{document}